\documentclass[showpacs,superscriptaddress, preprint,aps]{revtex4}
\usepackage{graphicx}% Include figure files
\usepackage{dcolumn}% Align table columns on decimal point
\usepackage{bm}% bold math
\usepackage{pstricks}
\usepackage{epsfig}
\begin{document}
%\preprint{APS/MWNT}

\title{Pressure induced re-emergence of superconductivity in superconducting topological insulator Sr$_{0.065}$Bi$_{2}$Se$_3$}

\author{Yonghui Zhou}
\affiliation{High Magnetic Field Laboratory, Chinese Academy of Sciences and University of Science and Technology of China, Hefei 230026, China}
\affiliation{Key Laboratory of Materials Physics, Institute of Solid State Physics, Chinese Academy of Sciences, Hefei 230031, China}

\author{Xuliang Chen}
\affiliation{High Magnetic Field Laboratory, Chinese Academy of Sciences and University of Science and Technology of China, Hefei 230026, China}
\affiliation{Key Laboratory of Materials Physics, Institute of Solid State Physics, Chinese Academy of Sciences, Hefei 230031, China}

\author{Ranran Zhang}
\affiliation{High Magnetic Field Laboratory, Chinese Academy of Sciences and University of Science and Technology of China, Hefei 230026, China}

\author{Jifeng Shao}
\affiliation{High Magnetic Field Laboratory, Chinese Academy of Sciences and University of Science and Technology of China, Hefei 230026, China}

\author{Xuefei Wang}
\affiliation{High Magnetic Field Laboratory, Chinese Academy of Sciences and University of Science and Technology of China, Hefei 230026, China}
\affiliation{Key Laboratory of Materials Physics, Institute of Solid State Physics, Chinese Academy of Sciences, Hefei 230031, China}

\author{Chao An}
\affiliation{High Magnetic Field Laboratory, Chinese Academy of Sciences and University of Science and Technology of China, Hefei 230026, China}
\affiliation{Key Laboratory of Materials Physics, Institute of Solid State Physics, Chinese Academy of Sciences, Hefei 230031, China}

\author{Ying Zhou}
\affiliation{High Magnetic Field Laboratory, Chinese Academy of Sciences and University of Science and Technology of China, Hefei 230026, China}
\affiliation{Key Laboratory of Materials Physics, Institute of Solid State Physics, Chinese Academy of Sciences, Hefei 230031, China}

\author{Changyong Park}
\affiliation{High Pressure Collaborative Access Team, Geophysical Laboratory, Carnegie Institution of Washington, Argonne, Illinois 60439, USA}

\author{Wei Tong}
\affiliation{High Magnetic Field Laboratory, Chinese Academy of Sciences and University of Science and Technology of China, Hefei 230026, China}

\author{Li Pi}
\affiliation{High Magnetic Field Laboratory, Chinese Academy of Sciences and University of Science and Technology of China, Hefei 230026, China}
\affiliation{Collaborative Innovation Center of Advanced Microstructures, Nanjing, 210093, China}

\author{Zhaorong Yang}
\thanks{zryang@issp.ac.cn}
\affiliation{High Magnetic Field Laboratory, Chinese Academy of Sciences and University of Science and Technology of China, Hefei 230026, China}
\affiliation{Key Laboratory of Materials Physics, Institute of Solid State Physics, Chinese Academy of Sciences, Hefei 230031, China}
\affiliation{Collaborative Innovation Center of Advanced Microstructures, Nanjing, 210093, China}

\author{Changjin Zhang}
\thanks{zcjin@ustc.edu.cn}
\affiliation{High Magnetic Field Laboratory, Chinese Academy of Sciences and University of Science and Technology of China, Hefei 230026, China}
\affiliation{Collaborative Innovation Center of Advanced Microstructures, Nanjing, 210093, China}

\author{Yuheng Zhang}
\affiliation{High Magnetic Field Laboratory, Chinese Academy of Sciences and University of Science and Technology of China, Hefei 230026, China}
\affiliation{Collaborative Innovation Center of Advanced Microstructures, Nanjing, 210093, China}

 \begin{abstract}
The recent-discovered Sr$_x$Bi$_2$Se$_3$ superconductor provides an alternative and ideal material base for investigating possible topological
superconductivity.
Here, we report that in Sr$_{0.065}$Bi$_{2}$Se$_3$, the ambient superconducting phase is gradually depressed upon
the application of external pressure.
At high pressure, a second superconducting phase emerges at above 6 GPa, with a maximum $T_c$ value of $\sim$8.3 K.
The joint investigations of the high-pressure synchrotron x-ray diffraction and electrical transport
properties reveal that the re-emergence of superconductivity in Sr$_{0.065}$Bi$_{2}$Se$_3$ is closely related to the structural phase transition
from ambient rhombohedral phase to high-pressure monoclinic phase around 6 GPa, and further to another high-pressure tetragonal phase above 25 GPa.
\end{abstract}

\pacs{74.70.Dd, 74.62.Fj, 74.25.Op, 75.47.-m}
\maketitle

Topological insulators and topological superconductors are new states of quantum matter which cannot be adiabatically connected to
conventional insulators and superconductors~\cite{Kane2005,Bernevig2006,Fuliang2008,Qixl2011}.
Topological insulators are characterized by a full insulating gap in the bulk and gapless edge or surface states which are protected
by time-reversal symmetry. In recent years, topological insulating materials have been theoretically predicted and experimentally
observed in a variety of systems, including HgTe quantum wells, BiSb alloys, TlBiSe$_2$ and Bi$_2$Te$_3$ (Bi$_2$Se$_3$, Sb$_2$Te$_3$) single
crystals~\cite{Konig2007,Hsieh2008,Zhanghj2009}.

The concept of topological insulator can also be applied to superconductor, due to the direct analogy between
topological band theory and superconductivity. Of major interest in the field of topological superconductivity is the realization of
Majorana fermions, that are predicted to exist as protected bound states in topological superconductors~\cite{Fuliang2008}.
In recent years, tremendous efforts have been put in order to realize a real topological superconducting
material~\cite{Hor2010,Kriener2011,Jincq2011,Xueqk2012,Zareapour2012,Xujp2014}. Among them, the Cu-intercalated Cu$_x$Bi$_2$Se$_3$
has attracted much attention, because large-size bulk superconducting single crystals can be obtained. Various experimental work as well
as theoretical analyses have been done, in order to reveal the novel physical properties in the Cu$_x$Bi$_2$Se$_3$
system~\cite{Fuliang2010,Wray2010,Sasaki2011,Lawson2012,Fuliang2012,Levy2013,Wanxg2014}. However, whether or not the Cu$_x$Bi$_2$Se$_3$
is a topological superconductor is still controversial. Recently, we found that an alternative
compound, Sr$_x$Bi$_2$Se$_3$, exhibits superconductivity with high superconducting volume fraction~\cite{Liuzh2015}.
The following investigations have revealed that the atomic position of Sr in Sr$_x$Bi$_2$Se$_3$ is
completely different from that of Cu in Cu$_x$Bi$_2$Se$_3$. That is, the copper atoms are intercalated in the weakly van der Waals bonded
Se-Se layers, while most of the intercalated strontium atoms are located in the Se-Bi-Se-Bi-Se quintuple layer~\cite{Qiand2015}.
Furthermore, the Sr$_x$Bi$_2$Se$_3$ compounds exhibit well-separated topological surface
state from the bulk bands~\cite{Liuzh2015,Qiand2015,Shruti2015,Neupane2015}. These facts suggest that the Sr$_x$Bi$_2$Se$_3$ compound
exhibits interesting physical phenomena and could serve as an important material base for the investigation of topological
superconducting-related properties.

In this work, we perform a systematic study on the electronic transport properties and the structural evolution of
Sr$_{0.065}$Bi$_{2}$Se$_3$ under high pressure.
We find that the ambient superconducting phase in Sr$_{0.065}$Bi$_{2}$Se$_3$ is gradually depressed with the application of pressure.
Noticeably, a pressure-induced superconducting phase emerges at high pressure. It is found that
re-emergence of superconductivity and the structural phase transition in Sr$_{0.065}$Bi$_{2}$Se$_3$ are quite comparable with
the high-pressure induced superconductivity in the Bi$_{2}$Se$_3$ pristine topological insulator, despite of the fact that
the Sr$_{0.065}$Bi$_{2}$Se$_3$ is a superconductor at ambient pressure.

Single crystal of Sr$_{0.065}$Bi$_{2}$Se$_3$ with typical dimensions of 3$\times$3$\times$0.5 mm$^3$ used in this work has been
reported previously~\cite{Liuzh2015}. High-pressure resistance measurements were conducted in a screw-pressure-type diamond anvil cell (DAC).
Diamond anvils of 200 $\mu$m culets and T301 stainless-steel gasket covered with a mixture of epoxy and fine cubic boron nitride (c-BN) powder
were used for high-pressure transport measurements. The four-probe method was applied in the $ab$-plane of single crystal with typical
dimension of 90$\times$40$\times$10 $\mu$m$^3$. The magnetoresistance experiments under high-pressure were performed on the Cell5
Water-Cooling Magnet of High Magnetic Field Laboratory of Chinese Academy of Sciences.
The measurements were done using a field-sweeping method at fixed temperature. The maximum magnetic field is 33 Tesla.
High-pressure synchrotron radiation x-ray diffraction measurements were performed at 16BMD~\cite{Park2015}, HPCAT,
Advanced Photon Source, Argonne National Laboratory. The as-grown single crystals were ground into fine powder for the x-ray diffraction
experiments with a wavelength of 0.4246 \AA. A rhenium gasket was pre-indented with a thickness of 40 $\mu$m. Then a 120 $\times$m hole
was drilled by the laser micro-machining system at HPCAT~\cite{Hrubiak2015}, which was served as a sample chamber.
A pre-pressed powder sheet with typical size of 30 $\mu$m$\times$30 $\mu$m$\times$15 $\mu$m was loaded into the chamber together
with a ruby ball and silicone oil served as pressure marker and pressure transmitting medium, respectively.
A two-dimensional area detector Mar345 was used to collect the powder diffraction patterns.
The Dioptas~\cite{Prescher2015} and Rietica~\cite{Howard1988} programs were employed for the image integrations and the XRD profile refinements, respectively.
Le Bail method was used to extract the lattice parameters. The pressure in the cell was measured at room temperature
with an offline ruby system at HPCAT.

Figure 1(a) gives the evolution of resistivity as a function of temperature for the Sr$_{0.065}$Bi$_{2}$Se$_3$ sample
at relatively low pressure ($P$$\leq$3 GPa).
Two distinct features can be found in this pressure region. One is that the normal state resistivity increases with increasing
pressure. The other is that the superconductivity is gradually depressed. If we compare the response of superconductivity of Sr$_{0.065}$Bi$_{2}$Se$_3$
with that of Cu$_{0.3}$Bi$_{2}$Se$_3$~\cite{Bay2012}, we find that
the depression of superconductivity in Sr$_{0.065}$Bi$_{2}$Se$_3$ sample is much faster than that in Cu$_{0.3}$Bi$_{2}$Se$_3$.
The depression of superconductivity in Sr$_{0.065}$Bi$_{2}$Se$_3$ can be qualitatively explained according to a
simple model for a low carrier density superconductor where $T_c$$\sim$$\Theta$$_D$exp[-1/$N$(0)$V_0$], with $\Theta$$_D$ the Debye temperature,
$N$(0)$\sim$$m^*$$n^{1/3}$ the density of states (with $m^*$ the effective
mass) and $V_0$ the effective interaction parameter~\cite{Bay2012}. In Sr$_{0.065}$Bi$_{2}$Se$_3$, the reduction of charge carrier density $n$
under pressure is apparent: The temperature dependence of resistivity gradually loses its metallic character with increasing pressure and the
$\rho$ (300 K) value exhibits an increase by a factor of $>$2 when the applied pressure is increased from 0.3 GPa to 2.7 GPa.

Figure 1(b) shows the temperature dependence of resistivity for the Sr$_{0.065}$Bi$_{2}$Se$_3$ sample with 2.7 GPa$\leq$$P$$\leq$40 GPa.
It is interesting to notice that the normal state resistivity again exhibits metallic-like feature, indicating an increase of
charge carrier density at high pressure. The normal state
resistivity decreases continuously with increasing pressure, suggesting a successive increase of charge carrier density at high
pressure region. A striking phenomenon is that the Sr$_{0.065}$Bi$_{2}$Se$_3$ sample exhibits
a re-emergent superconductivity when $P$$\geq$6 GPa. It should be mentioned that this re-emergent superconductivity has not been observed in
Cu$_{x}$Bi$_{2}$Se$_3$ compound. In order to see the re-emergent superconductivity more clearly, we plot in Fig. 1(c) an enlarged view
near the transition temperature.
The $T_c^{onset}$ value at 6.0 GPa is about 3.6 K. And the $T_c^{onset}$ slightly decreases with increasing pressure
when 6 GPa$\leq$$P$$\leq$11.5 GPa. However, when $P$$\geq$14 GPa, the $T_c^{onset}$ value is drastically increased, reaching to a maximum of
$\sim$8.3 K. With further increasing pressure, the $T_c^{onset}$ is slightly decreased.
Nevertheless, the re-emergent superconductivity of Sr$_{0.065}$Bi$_{2}$Se$_3$ is quite robust under high pressure. The superconductivity occurs even
at the highest achieved pressure of 80 GPa(Fig. 1(d)). We also notice that
in iron chalcogenide superconductors, such as Tl$_{0.6}$Rb$_{0.4}$Fe$_{1.67}$Se$_2$, K$_{0.8}$Fe$_{1.7}$Se$_2$ and K$_{0.8}$Fe$_{1.78}$Se$_2$,
the re-emergent superconductivity with higher transition temperature has been reported~\cite{Sunll2012}. However, in iron chalcogenides,
the re-emergent superconductivity occurs in a much narrower pressure range comparing to the present case.

In order to tentatively estimate the nature of the re-emergent superconductivity, we analyze the evolution of the upper critical field $H_{c2}$($T$)
at low temperature.
Figure 2(a) shows the response of the re-emergent superconductivity on external magnetic field. The applied pressure is fixed at 19.5 GPa.
It is found that the superconducting transition temperature is monotonously decreased with increasing magnetic field.
One striking feature is that both with and without magnetic field, the superconducting transition is rather
sharp, meaning the occurrence of bulk and homogenous superconductivity. It can be also found that the $\rho$$\sim$$T$ curves are almost
parallel to each other, suggesting that the flux creep effects can be completely ignored in the vortex dynamics of the re-emergent superconductivity.
Figure 2(b) plots the dependence of the reduced critical field, $h$$^*$($T$)=[$H_{c2}$($T$)/$T_c$]/[$d$$H_{c2}$($T$)/$d$$T$]$|$$_{T=T_c}$,
on the normalized temperature $t$=$T/T_c$. Here the $H_{c2}$($T$) value is defined from the
resistance criterion of $R_{cri}$ = 90\%$R_n$ ($R_n$ is the normal state resistance).
The experimental data is compared to models for orbitally limited $s$-wave
and spin-triplet $p$-wave superconductors.
It can be seen that the experimental $h$$^*$($T$) deviates significantly from the expected orbital-limited behavior predicted by the
Werthamer-Helfand-Hohenberg (WHH) theory for an $s$-wave superconductor ($h$$^*$(0)$\simeq$0.72)~\cite{Werthamer1966}.
Noticeably, the $h$$^*$($T$)
data is very close to the $h$$^*$($T$)$\sim$$t$ curve expected from a $p$-wave superconductor ($h$$^*$(0)$\simeq$0.85)~\cite{Scharnberg1980}.
We also notice that the better satisfying of the $h$$^*$($T$) data to a $p$-wave model rather than a $s$-wave model has been reported in
pressure-induced superconductivity in Bi$_{2}$Se$_3$ compound~\cite{Kirshenbaum2013}, pointing to unconventional superconductivity
in the Sr$_x$Bi$_{2}$Se$_3$ system.

It is instructive to investigate the structural symmetry at high pressure.
Thus we conduct the high-pressure synchrotron x-ray diffraction (XRD) study on the Sr$_{0.065}$Bi$_{2}$Se$_3$ sample up to 41.1 GPa.
In Fig. 3, the powder diffraction patterns and the lattice parameters under high pressure are presented in detail.
Selected XRD profile refinements are shown in Supplementary Fig. 3. It is found that the ambient rhombohedra phase
($R$-3$m$) can remain in a single phase only when $P$$\leq$5.7 GPa, above which a monoclinic phase ($C$2/$m$) is involved
and coexists with the rhombohedra one. The low-pressure $R$-3$m$ phase disappears completely at 16.6 GPa
and the $C$2/$m$ phase persists up to 31.6 GPa. In the pressure range of 25.0-31.6 GPa, a mixture of the $C$2/$m$ phase
and the high pressure body-centered tetragonal phase ($I$4/$mmm$) shows up. Upon further compression beyond 31.6 GPa,
the $I$4/$mmm$ phase can be sustained alone up to the highest pressure achieved in the present experiment.

To obtain a comprehensive understanding of the pressure driven superconducting behavior, we show the isothermal equations of state
for the respective phases in Fig. 4. The isothermal equations of state are fitted by the third-order Birch-Murnaghan formula~\cite{Birch1947}
as indicated by the red solid lines in Fig. 4(a). With $B_0$' fixed as 4, the isothermal bulk modulus $B_0$ is estimated to be
$\sim$58(1), 85(1), and 116(2) GPa for the $R$-3$m$, $C$2/$m$, and $I$4/$mmm$ phases, respectively.
It is found that at the two progressive structural transition points
($R$-3$m$$\rightarrow$$C$2/$m$$\rightarrow$$I$4/$mmm$), the unit-cell volume per chemical formula (Sr$_{0.065}$Bi$_{2}$Se$_3$)
shrinks by about 4.8\% and 4.0\%, respectively. According to the structural symmetries under high pressure, the superconducting phase diagram
can be divided into two different regions as shown in Fig. 4(b): the pristine superconducting phase SC-I and the pressure-induced
superconducting phase SC-II at high pressure. In the first region, the superconductivity is gradually depressed with
increasing pressure. The $T_c$ value approaches to zero at an extrapolated pressure of about 1.1 GPa.
Note that within this pressure range, the crystal structure of Sr$_{0.065}$Bi$_{2}$Se$_3$ is in the rhombohedral $R$-3$m$ phase without
structural phase transition. With increasing pressure, the $C$2/$m$ phase is involved and coexists with the rhombohedral $R$-3$m$ phase.
Meanwhile, the re-emergent superconductivity occurs above 6 GPa. The superconducting critical temperature first experiences a tiny decrement
till about 12 GPa. Then the $R$-3$m$ phase disappears completely, leaving the $C$2/$m$ one stays alone. At the same time, the superconducting
transition temperature jumps abruptly and reaches rapidly a maximum value of $\sim$8.3 K,
which is followed by a slight decrease again. Upon further compression, the $C$2/$m$ to $I$4/$mmm$
structural phase transition occurs at around 25 GPa. As a matter of fact, previous high pressure studies on Bi$_{2}$Se$_3$ have revealed similar
structural transition from the ambient-pressure rhombohedral ($R$-3$m$) structure to a
lower-symmetry monoclinic ($C$2/$m$) structure near 10 GPa, and then to an unknown phase above 28 GPa~\cite{Kirshenbaum2013,Vilaplana2011}.
Thus the structural phase transition in Sr$_{0.065}$Bi$_{2}$Se$_3$ at high pressure is similar to that in Bi$_{2}$Se$_3$ pristine topological
insulator.

In conclusion, a novel pressure induced re-emergent superconductivity has been revealed in Sr$_x$Bi$_{2}$Se$_3$ superconducting topological
insulator. The resulting phase diagrams exhibit similar features to those of the high pressure response of Bi$_{2}$Se$_3$ pristine
topological insulator, including the role of structural transitions and the presence of unconventional superconductivity. The
analysis on the pressure-invariant $T_c$ suggests the unconventional nature of the superconductivity in Sr$_x$Bi$_{2}$Se$_3$,
which deserves further investigation.

This work was supported by the National Natural Science Foundation of China (Grant Nos. U1532267, U1332143, and 11574323).
A portion of this work was performed on the WM5 magnet of the High Magnetic Field Laboratory, Chinese Academy of Sciences.
Portions of this work were performed at HPCAT (Sector 16), Advanced Photon Source (APS), Argonne National Laboratory (ANL).
HPCAT operations are supported by DOE-NNSA under Award No. DE-NA0001974 and DOE-BES under Award No. DE-FG02-99ER45775,
with partial instrumentation funding by NSF. APS is supported by DOE-BES, under Contract No. DE-AC02-06CH11357.
Y. H. Zhou and X. L. Chen contributed equally to this work.

\begin{description}
\item  \bigskip FIGURE CAPTION
\end{description}

FIG. 1 (color online)  (a) Temperature-dependent $ab$-plane resistance for Sr$_{0.065}$Bi$_{2}$Se$_3$ at low pressure region.
(b) Temperature-dependent $ab$-plane resistance for Sr$_{0.065}$Bi$_{2}$Se$_3$ at high pressure region.
(c) An enlarged view of the re-emergent superconductivity near the transition temperature (Run 1).
(d) The temperature dependence of resistance for the Sr$_{0.065}$Bi$_{2}$Se$_3$ at Run 2, reaching a maximum pressure of 80 GPa.

\bigskip

FIG. 2 (color online) (a)  Temperature dependence of resistance under different magnetic fields up to 5.0 T. The applied pressure is fixed at 19.5 GPa.
(b) Temperature dependence of the reduced upper critical field $h^*$ (the red circles) and the fittings according to a $p$-wave polar
state and the $s$-wave clean limit, respectively. The $T_c$ at specific magnetic field is determined as 90\% drop of
the normal state resistance.

\bigskip

FIG. 3 (color online) (a) Synchrotron radiation X-ray diffraction patterns at various applied pressures.
For clarity, the backgrounds have been subtracted by using the Dioptas program.
(b) The refined lattice parameters $a$ (square), $b$ (diamond) and $c$ (circle) as a function of pressure.
Three structural phases, i. e., $R$$-$3$m$, $C$2/$m$, $I$4/$mmm$, with two overlap pressure regions are revealed upon compression.

\bigskip

FIG. 4 (color online) (a)  The compression data ($V$ versus $P$) was fitted by the three-order Birch-Murnaghan equation of state (solid red line),
which yields the bulk modulus 58(1), 85(1), 116(2) GPa for the $R$$-$3$m$, $C$2/$m$, and $I$4/$mmm$ phases, respectively.
(b) The pristine superconducting phase, SC-I, occurs in the ambient rhombohedral ($R$$-$3$m$) phase.
The re-emergent superconducting phase, SC-II, occurs in the monoclinic ($C$2/$m$) phase and in the tetragonal $I$4/$mmm$ phase.

\end{document}